\documentclass[apm,reprint,amsmath,amssymb]{revtex4-1}

\usepackage[latin1]{inputenc}
\usepackage{graphics}
\usepackage{color}
\usepackage{amssymb}
\usepackage{amsmath}
\usepackage{overpic}
\usepackage{epstopdf}
\usepackage{bm}
\usepackage{graphicx}
\usepackage{subfigure}
\usepackage{multirow}

\begin{document}

\title{Thermal conductivity of perovskite
KTaO$_3$ and PbTiO$_{3}$ from first principles}

\author{Yuhao Fu}
\author{David J. Singh}

\affiliation{Department of Physics and Astronomy, University of Missouri,
Columbia, MO 65211-7010, USA}

\date{\today}

\begin{abstract}
The low thermal conductivity of piezoelectric perovskites is a challenge
for high power transducer applications.
We report first principles calculations of the thermal conductivity of
ferroelectric PbTiO$_3$ and the cubic nearly ferroelectric perovskite
KTaO$_3$. The calculated thermal conductivity of PbTiO$_3$ is much
lower than that of KTaO$_3$
in accord with experiment. 
Analysis of the results shows that the reason for the low thermal conductivity
of PbTiO$_3$ is the presence of low frequency optical phonons
associated with the polar modes. These are less dispersive in PbTiO$_3$,
leading to a large three phonon scattering phase space.
These differences between the two materials
are associated with the $A$-site driven ferroelectricity
of PbTiO$_3$ in contrast to the $B$-site driven near ferroelectricity of
KTaO$_3$.
The results are discussed in the context of modification of the
thermal conductivity of electroactive materials.

\end{abstract}

\maketitle

\section{Introduction}

The low thermal conductivity of piezoelectric oxides, such as
Pb(Zr,Ti)O$_3$ (PZT) ceramics, limits the performance of high
power transducers for applications such as sonar and ultrasonics.
\cite{stulen,decarpigny,dubus,butler1}
It is also a constraint on emerging applications such as pyroelectric
energy harvesting. \cite{wang}
While this can be mitigated by careful design of the transducers,
it is desirable to develop understanding of the low thermal conductivity
of these materials, perhaps leading to modifications that can
improve the heat conduction.
Common PZT ceramics have ultralow thermal conductivities of
$\kappa_l \sim$2 W/mK or less at ambient temperature, decreasing with
temperature to the ferroelectric phase transition, and then
increasing sharply above the ferroelectric transition, \cite{kallaev}
implying an association between the ultralow thermal conductivity
and the ferroelectricity.
It is also important to note that newer high performance materials, 
such as
Pb(Mg$_{1/3}$Nb$_{2/3}$)O$_3$ - PbTiO$_3$ alloy (PMN-PT) and
Pb(Zn$_{1/3}$Nb$_{2/3}$)O$_3$ - PbTiO$_3$ alloy (PZN-PT) piezocrystals
have more strongly temperature dependent properties than PZT.
This means that local heating may lead to strong changes in local
dielectric and piezoelectric constants and resulting concentrations
of electric field and mechanical stress potentially resulting in breakdown
failure due to hotspots.

In other ferroelectric and near ferroelectric
materials, particularly rocksalt structure chalcogenides, the
interplay between polar transverse optical phonons and the longitudinal
acoustic (LA) branch has been implicated in strongly lowering the thermal
conductivity. \cite{delaire,romero,cwli,murphy,shulumba}
This is through a strong anharmonic coupling of the ferroelectric polar mode
with the longitudinal acoustic mode, as seen directly in the very strong
volume dependence of that mode, i.e. the high Gr\"uneissen parameter. In
addition, phase space considerations are important.
\cite{cwli,murphy,shulumba}

Similarly, in perovskite $AB$O$_3$
ferroelectrics, there is typically a very strong
pressure dependence to the polar mode and associated phase transitions,
\cite{samara,fornari,sani,jaouen}
indicating strong anharmonic coupling with longitudinal acoustic phonons
near the zone center, where the LA modes are compression waves.

This raises the questions of whether it is possible to have high thermal
conductivity in a high performance
perovskite piezoelectric, and if so how that might
be achieved. Specifically, if ferroelectricity or proximity to ferroelectricity
is the origin of the low thermal conductivity, and also is essential
to the piezoelectric performance, it could be that low thermal conductivity
is unavoidable.
On the other hand, the perovskite structure is considerably more complex
than the rocksalt structure, and mechanisms for ferroelectricity in
perovskites are similarly more complex, which may provide avenues for
increasing thermal conductivity that
are not available in the nearly ferroelectric rocksalt compounds.

Here we report detailed first principles
thermal conductivity calculations using
solution of the Boltzmann-Peierls equation based on calculated three
phonon anharmonic couplings for two prototypical materials with 
distinct behavior. These are PbTiO$_3$, which is a high polarization
ferroelectric that is the tetragonal end-member of the PZT solid solution,
and KTaO$_3$, which is a nearly ferroelectric perovskite with nearness
to ferroelectricity coming from a different mechanism than in PbTiO$_3$.
Specifically, ferroelectricity is regarded as $A$-site (Pb) driven in
PbTiO$_3$ but $B$-site (Ta) driven in the near ferroelectric KTaO$_3$.
\cite{ghita}
These two compounds have the advantages of representing different
behaviors and having rather different experimental thermal conductivities,
while retaining relatively simple five atom unit cells,
which facilitates calculations and analysis.
Additionally, there is high quality single crystal experimental 
thermal conductivity data available for these compounds.
\cite{tachibana}
We find that proximity to ferroelectric phase transitions is not
the only factor in the thermal conductivity, but rather that the
distinct phonon dispersions of $A$-site vs. $B$-site driven
perovskites play a key role, suggesting possible avenues for
improving the thermal conductivities.

\section{Computational Methods}

\begin{figure}[h]
\includegraphics[width=\columnwidth]{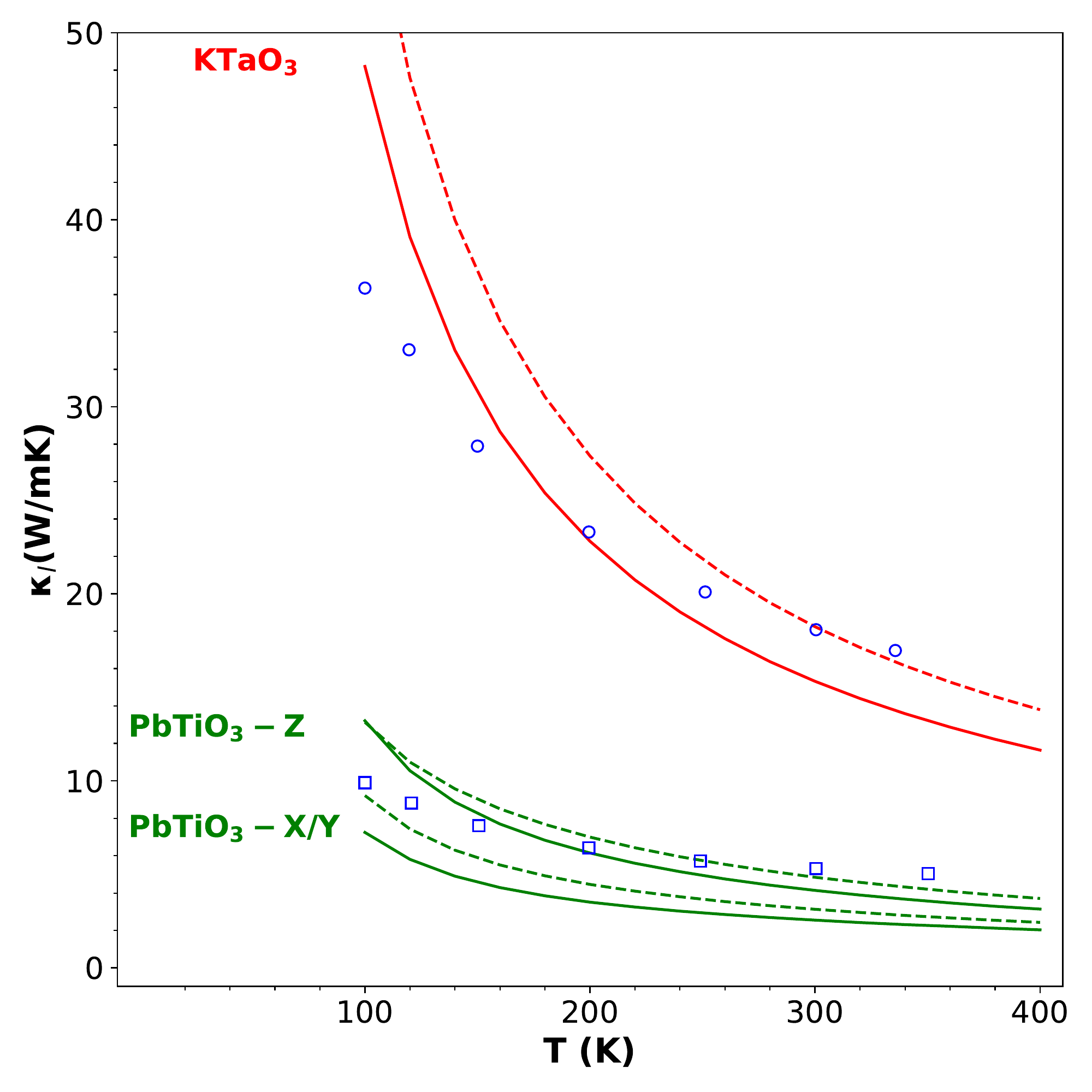}
\centering
\caption{Calculated temperature dependence of lattice thermal
conductivity $\kappa_{l}$ (in W/mK)
of KTaO$_{3}$ and PbTiO$_{3}$.
The solid lines indicate the $\kappa_{l}$ for calculations
with experimental lattice parameters,
while the dashed lines indicate the $\kappa_{l}$ with full
relaxation including the lattice parameters.
The experimential $\kappa_{l}$ for KTaO$_{3}$ (open circles)
and PbTiO$_{3}$ (open squares) are taken from
Tachibana et al. \cite{tachibana}.}
\label{thermalConductivity}
\end{figure}

We performed first principles calculations of $\kappa_{l}$
for perovskite KTaO$_3$ and PbTiO$_3$ in their ground state structures
by iteratively solving the linearized Boltzmann-Peierls
transport equation of phonons with the ShengBTE package. \cite{li2014}
Properties of
ferroelectrics are often highly sensitive to the lattice parameters.
We used the experimental values of the ambient temperature lattice
parameters,
\cite{ktao3, pbtio3} 
and also did
calculations with fully relaxed lattice parameters.
The internal atomic atomic positions were relaxed using total energy
minimization.
The experimental and calculated lattice parameters are
$a$=3.9883 \AA, and $a$=3.9669 \AA, respectively, for KTaO$_3$ and
$a$=3.904 \AA, $c$=4.1575 \AA, and
$a$=3.865 \AA, $c$=4.036 \AA, respectively, for PbTiO$_3$.
The interatomic force constants (IFCs)
were obtained from density functional theory (DFT) calculations with the
projector augmented wave (PAW) method \cite{PhysRevB.50.17953},
as implemented in the VASP code \cite{PhysRevB.54.11169}.
The 3p and 4s (K), 2p, 3d and 4s (Ti), 5p, 5d and 6s (Ta),
5d, 6s and 6p (Pb), 2s and 2p (O) electrons
were treated explicitly as valence electrons.
We used the local density approximation (LDA)
for the exchange-correlation functional.  

The kinetic energy cutoffs for the plane-wave basis set were set to 520 eV.
The k-point meshes with grid spacing of 2$\pi\times$0.03 \AA$^{-1}$
were used for electronic Brillouin zone integration.
Structural optimizations were done
with a tolerance on residual forces of 10$^{-4}$ eV/\AA.
The harmonic and third-order anharmonic IFCs were calculated by using the
real-space supercell approach \cite{li2014, phonopy} with
a 3$\times$3$\times$3 $k$-point mesh defining the supercells.
The longitudinal optical - transverse optical (LO-TO) phonon splitting is
large in these compounds, and was included using the dielectric
constant and Born effective charges calculated from linear response
density functional perturbation theory as implemented in the VASP code.
Phonon momenta
$q$-meshes of 15$\times$15$\times$15
were used in solving the transport equation, which was sufficient to
converge $\kappa_{l}$ to a very high accuracy with respect to this parameter.

\section{Results and Discussion}

\begin{figure}[h]
\includegraphics[width=\columnwidth]{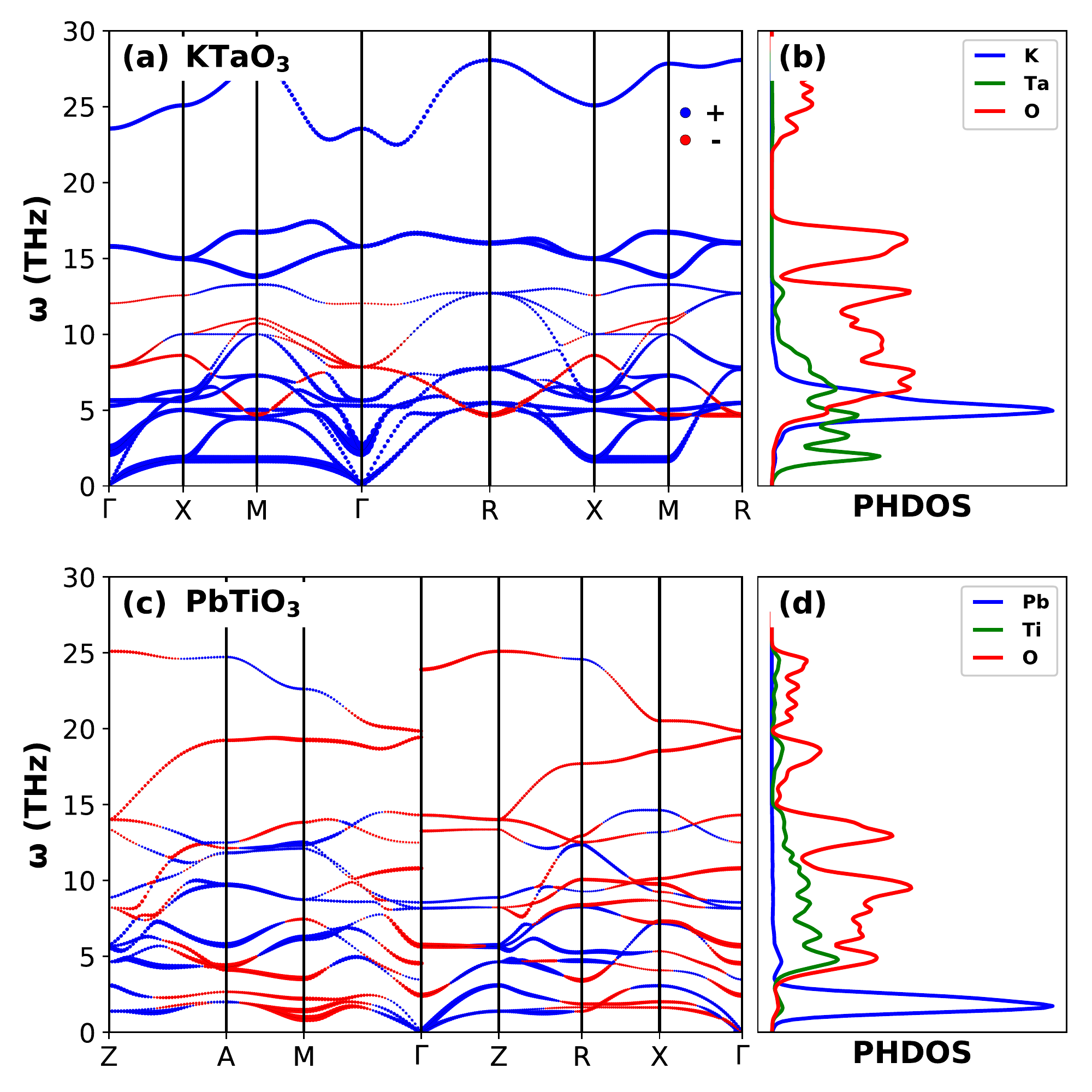}
\centering
\caption{(a, c) Calculated phonon dispersions
of KTaO$_{3}$ and PbTiO$_{3}$.
The symbol sizes give the Gr\"uneisen parameter ($\gamma$) 
of each phonon mode. The colors, blue (+) and red (-) indicate positive
and negative $\gamma$.
The $\gamma$ values of the lowest frequency phonons
at M, X and R for KTaO$_3$ are 11.33, 12.03 and -3.19,
respectively. For PbTiO$_3$,
the $\gamma$ values of the lowest frequency phonons
at Z, A and M are 0.87, 0.20 and -14.45, respectively.
The projected phonon densities of states (PHDOS)
are shown in (b) and (d).}
\label{phonon}
\end{figure}

\begin{table}[h]
\caption{Calculated lowest phonon frequencies at symmetry points,
compared with experimental data.
\cite{PhysRev.157.396, PhysRevB.5.1886, PhysRevB.48.10160, PhysRevB.73.140101}
}
\centering
\begin{tabular}{lccc}
\hline\hline
                                   &                     & Calc. (THz) & Exp. (THz, 300K) \\
\hline
\textbf{KTaO$_{3}$} & $\Gamma$   & 2.34          & 2.56$^{a}$ \\
                                   &                      & 5.63          & 6.24$^{a}$ \\
                                   &                      & 7.84          & 8.58$^{a}$ \\
                                   & M                  & 1.76          & 1.82$^{b}$ \\
                                   &                      & 5.02          & 5.27$^{a}$ \\
                                   &                      & 10.00        & 10.28$^{a}$ \\
                                   & X                   & 1.76          & 1.90$^{a}$  \\
                                   &                      & 5.01          & 5.22$^{a}$ \\
                                   &                      & 5.85          & 5.99$^{a}$ \\
                                   & R                   & 4.66          & 4.90$^{a}$  \\
                                   &                      & 5.48          & 5.76$^{a}$ \\
                                   &                      & 7.69          & 7.63$^{a}$ \\
\hline
\textbf{PbTiO$_{3}$} & $\Gamma$   & 2.45          & 2.61$^{c}$  \\
                                    &                     & 3.48          & 3.84$^{c}$ \\
                                    &                     & 4.55          & 4.47$^{c}$ \\
                                   & X                  & 1.64          & 1.62$^{d}$  \\
                                   &                      & 2.01          & 1.97$^{d}$ \\
                                   &                      & 3.07          & 2.96$^{d}$ \\
\hline
$^{a}$Reference\cite{PhysRev.157.396}. & & & \\
$^{b}$Reference\cite{PhysRevB.5.1886}. & & & \\
$^{c}$Reference\cite{PhysRevB.48.10160}. & & & \\
$^{d}$Reference\cite{PhysRevB.73.140101}. & & & \\
\end{tabular}
\label{frequencies}
\end{table}

Our calculated thermal conductivity is shown in Fig. \ref{thermalConductivity}
along with comparison to experimental data in single crystals.
\cite{tachibana}
As seen, there is a very good agreement of the magnitude of the
thermal conductivity with experiment, and in particular,
the fact that KTaO$_3$ has a substantially higher thermal conductivity
than PbTiO$_3$ is reproduced. The temperature dependence is, however,
somewhat stronger in the calculated results than in experiment.
One possible reason for this discrepancy may be that the 
low frequency phonons are temperature dependent
in these materials, while we use the harmonic zero temperature
frequencies in the calculations presented here.
In any case,
it is clear that the calculations do capture the large difference
between PbTiO$_3$ and KTaO$_3$, which we discuss below, starting with the
phonons.

\begin{figure}[h]
\includegraphics[width=\columnwidth]{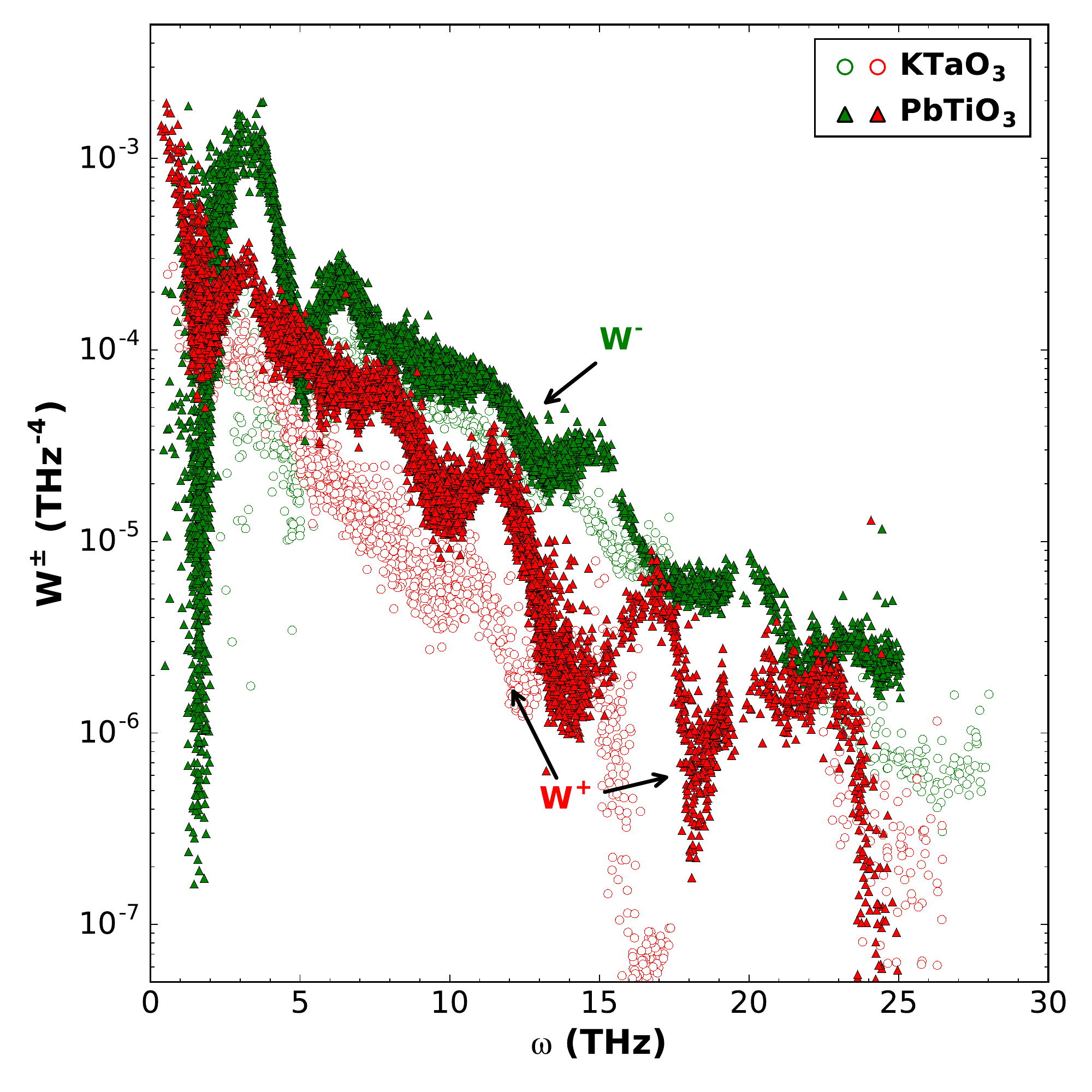}
\centering
\caption{Calculated three-phonon anharmonic scattering phase space
(W$^{\pm}$) of KTaO$_{3}$ and PbTiO$_{3}$.
The +(-) sign represents three-phonon absorption (emission) phase space.
Note the log scale.}
\label{w}
\end{figure}

The phonon dispersions and projected phonon densities of states
of the two materials are given in Fig. \ref{phonon}. 
Gr\"uneissen parameters, $\gamma$ are indicated on the dispersions.
Table \ref{frequencies} gives a
comparison between calculated low frequency phonons and 
available experimental data.
\cite{PhysRev.157.396, PhysRevB.5.1886, PhysRevB.48.10160, PhysRevB.73.140101}
Overall, we find very good agreement between our calculated
phonon frequencies and experimental data.
It should, however, be mentioned that the lowest 
mode at $\Gamma$ in KTaO$_3$ is the soft mode, which is strongly temperature
dependent; \cite{glinsek}
for this mode,
agreement with experiments done at 300 K is misleading; its
frequency is overestimated in our calculations
if one considers the temperature dependence.

In any case, as seen in the phonon dispersions,
the two materials are very different.
KTaO$_3$ has a low frequency two fold degenerate transverse optic
mode at the zone center, as expected.
This mode has a sizable positive
Gr\"uneissen parameter, implying strong anharmonic
coupling to the longitudinal acoustic mode.
It is associated with $B$-site (Ta)
off-centering in its octahedral cage, along
with smaller $A$-site displacements.
\cite{singh-ktao3}
This mode anti-crosses with the acoustic
branches and remains low frequency in sheets around the $\Gamma$-X-M planes,
i.e. the planes defined by $k_x$=0, $k_y$=0 or $k_z$=0.
These phonons are not, however, low frequency elsewhere in the Brillouin
zone, as seen from the dispersions along $\Gamma$-R, X-R and M-R.
This sheet like structure of the regions of the zone
where the $B$-site off-centering phonons are low frequency corresponds
to the chain-like displacements noted in early diffuse scattering
experiments for the $B$-site driven ferroelectrics,
KNbO$_3$ and BaTiO$_3$. \cite{comes,yu}

The phonons of PbTiO$_3$ are very different,
characteristic of $A$-site driven
materials, \cite{ghita}
and consistent with prior calculations and experiment.
\cite{ghosez,perry,tomeno}
There is a prominent
low frequency peak in the phonon density of states with strong $A$-site (Pb)
participation. This comes
off-centering polar modes,
which are associated with the ferroelectricity. These have primary Pb
character, and smaller cooperative Ti displacements, as typical for
$A$-site driven ferroelectrics. \cite{ghita}
Also, different from KTaO$_3$ there are additional low frequency
modes at the zone corners, M, corresponding to rotation of the TiO$_6$
octahedra. These modes have sizable negative Gr\"uneisen parameter,
meaning that they soften with lattice compression, characteristic
of tilt modes in titanate perovskites. \cite{fornari,islam,xiang}
Note that at the zone center the polar modes associated with Pb
off-centering also have negative Gr\"uneisen parameter. This is because
pressure suppresses ferroelectricity, which for the already ferroelectric
perovskite PbTiO$_3$ means pressure moves the system towards the ferroelectric
instability, and therefore softens the polar modes.
Also, importantly, the Pb off-centering modes associated with ferroelectricity
do not have the chain structure that is present in $B$-site driven perovskites.
Instead they have low frequency throughout the Brillouin zone.
These predominantly Pb modes underlie the low frequency peak in the
phonon density of states.
The key finding is that the main difference between the dynamics of PbTiO$_3$
and KNbO$_3$ is not in the anharmonicity, but is in the phonon dispersions.

\begin{figure}[h]
\includegraphics[width=\columnwidth]{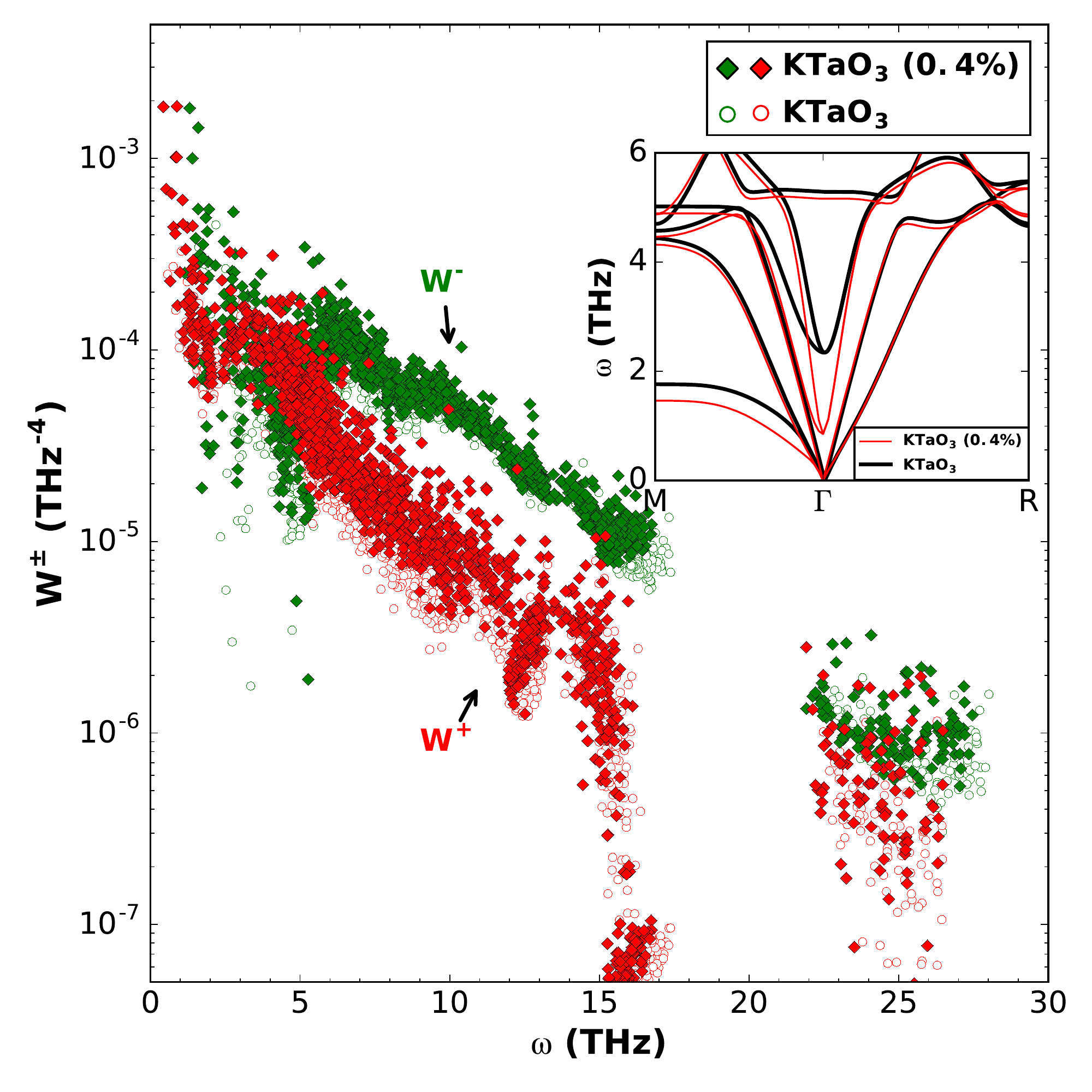}
\centering
\caption{Calculated three-phonon anharmonic scattering phase space
(W$^{\pm}$) of KTaO$_{3}$ and KTaO$_{3}$
(0.4\% of expansion in lattice parameters).
The +(-) sign represents three-phonon absorption (emission) process.}
\label{w4kto}
\end{figure}

This is reflected in the scattering phase space. Fig. \ref{w} shows the
calculated three-phonon anharmonic absorption and emission phase space for the
two compounds at the experimental lattice parameter. As seen, the scattering
phase space is very much larger for PbTiO$_3$ than for KNbO$_3$, including
importantly the low frequency region below $\sim$5 THz that is important
for heat conduction.
This explains the very low thermal conductivity of PbTiO$_3$
relative to KTaO$_3$. In particular, the fact that in perovskite structure,
the $A$-site motions are less correlated than the $B$-site motions in
$B$-site driven materials, leads to flat low frequency branches associated
with the polar mode, extending throughout the Brillouin zone. This
provides a large phase space for anharmonic phonon scattering and thus
low thermal conductivity.

We now turn to the issue of proximity to the ferroelectric instability.
Fig. \ref{w4kto}
shows the scattering phase
space for KTaO$_3$ for two different lattice parameters,
while the inset shows the corresponding change in phonon dispersion.
In KTaO$_3$ the volume expansion lowers the polar mode frequency bringing
the material much closer to the ferroelectric instability. This is reflected
in an increase in the scattering phase space at low frequency. However, the
changes are modest compared to the large difference between PbTiO$_3$ and
KTaO$_3$.

The resulting reduction in calculated thermal conductivity at 300 K
is from 15.3 W/mK to 11.1 W/mK, even though with the volume expansion the
material is much closer to the ferroelectric instability as seen in the
phonon dispersion.
Thus, while proximity to ferroelectricity is important, it is not
the main issue in the low thermal conductivity of PbTiO$_3$, since the
largest contribution to the scattering is not from the region of the zone
very close to the $\Gamma$ point.

\section{Summary and Conclusions}

Thermal conductivity calculations show that the low thermal conductivity
of PbTiO$_3$ in relation to KTaO$_3$ reflects the difference between
$A$-site driven and $B$-site driven polar perovskite materials, and
not the proximity to ferroelectricity. The implication is that it may be
possible to find chemical modifications that maintain the electromechanical
properties of piezoelectric perovskite ferroelectrics, while increasing
the thermal conductivity. This understanding may also be of use in related
materials classes, for example,
in perovskite thermoelectrics related to SrTiO$_3$, \cite{sun}
where reduction of the
thermal conductivity is desired,
as is typically the case for thermoelectrics. \cite{delaire,rogl}

One strategy suggested by the present results
may be to find chemical ways of increasing the coupling between
$A$-site and  $B$-site displacements. The rationale
is that it is the correlation
between $B$-site displacements that provides the dispersion of the low
frequency polar modes in perovskites, as in the chain-correlations of
KNbO$_3$. This correlation between $B$-site displacements is usually
understood as a consequence of covalency between
the $B$-site and O, resulting in over- and
under-bonding of O ions if the displacements
are not coherent. \cite{grinberg,liu}

It will therefore be of interest to explore perovskite solid
solutions based on $A$-site driven electroactive materials to find
systems with enhanced coupling between $A$-site and $B$-site displacements
and/or enhanced $B$-site -- O covalency, to determine if thermal
conductivity can be enhanced while retaining desirable piezoelectric
properties.
This may also lead to better control of thermal issues in applications
and perhaps better control of breakdown issues in pieozocrystals.

\acknowledgements

We are grateful for a helpful discussion with Lane Martin.

\bibliography{ferroelectric}

\end{document}